\documentclass{osa-article}

\journal{oe}
\usepackage{lineno}
\usepackage{mathtools}
\usepackage{amsmath}
\usepackage{breqn}

\begin{document}

\title{Deep sub-wavelength localization of light and sound in dielectric resonators}

\author{Alkim Bozkurt\authormark{1,2}, Chaitali Joshi\authormark{1,2}, and \mbox{Mohammad Mirhosseini \authormark{1,2,*}}}

\address{\authormark{1}The Gordon and Betty Moore Laboratory of Engineering, California Institute of Technology, Pasadena, California 91125\\
\authormark{2}Institute for Quantum Information and Matter, California Institute of Technology, Pasadena, California 91125}

\email{\authormark{*}mohmir@caltech.edu} %% email address is required



\begin{abstract}
Optomechanical crystals provide coupling between phonons and photons by confining them to commensurate wavelength-scale dimensions. We present a new concept for designing optomechanical crystals capable of achieving unprecedented coupling rates by confining optical and mechanical waves to deep sub-wavelength dimensions. Our design is based on a dielectric bowtie unit cell with an effective optical/mechanical mode volume of  $7.6 \times 10^{-3} {(\lambda/n_{\textrm{Si}})}^3$/$ 1.2 \times 10^{-3} {\lambda_{\textrm{mech}}}^3$. We present results from numerical modeling, indicating a single-photon optomechanical coupling of 2.2 MHz with experimentally viable parameters. Monte Carlo simulations are used to demonstrate the design's robustness against fabrication disorder.
\end{abstract}

%%%%%%%%%%%%%%%%%%%%%%%%%%  body  %%%%%%%%%%%%%%%%%%%%%%%%%%
\section{Introduction}
%Adherence to the specifications listed in this template is essential for efficient review and publication of submissions. Proper reference format is especially important (see Section \ref{sec:refs}).

The interaction between optical waves and displacement in mechanical structures has been studied extensively in the field of cavity optomechanics. While the initial work in this area focused on observing signatures of radiation pressure in the strongly driven macroscopic systems, later experiments enabled coherent control of nano-mechanical modes in the quantum regime \cite{Aspelmeyer:2014ce}. Optomechanical crystals (OMCs), made from periodic patterns of photonic and phononic crystals, have been particularly successful in this avenue. The strong interaction rates caused by co-localization of optical and mechanical waves in these devices has enabled a range of demonstrations such as quantum ground-state cooling \cite{Chan:2011dy}, photon-phonon entanglement \cite{10.1038/nature16536}, coherent control of solid-state atomic defects \cite{machielse2019}, and transduction of microwave to optical photons \cite{mirhosseini2020a}.

Previous efforts on designing optomechanical crystals have focused on maximizing the interaction rate while reducing the mechanical and optical sources of dissipation. These requirements can be met in micromachined silicon structures, where the large dielectric index and photo-elastic coefficients result in single-photon coupling rates above $1~\text{MHz}$, together with optical quality factors in the range $10^5-10^6$ \cite{chan2012}. Further, GHz-frequency silicon mechanical modes are shown to achieve ultralong decay times (exceeding 1 s) at cryogenic temperatures \cite{10.1126/science.abc7312}. Despite these attractive features, current state-of-the-art devices suffer from small phonon-photon conversion efficiencies because of their limited (optical) power handling capability at cold temperatures \cite{meenehan2015}. In light of these limitations, devices with enhanced single-photon optomechanical coupling are desired. However, obtaining significant improvement with existing design procedures appear challenging as previous results have been obtained through an extensive numerical optimization procedure \cite{chan2012}.

Here, we pursue an alternative design strategy that is motivated by recent development of ultra-small mode volume photonic crystal cavities \cite{hu2016a,hu2018a}. As we show, applying this concept to the design of dielectric optomechanical resonators leads to enhanced coupling rates by reducing the effective optical and mechanical mode volumes to the deep sub-wavelength regime. Using finite element method (FEM) simulations, we show that a single-photon optomechanical coupling rate of 2.2 MHz can be achieved with this design for experimentally viable parameters. We perform Monte Carlo simulations to model the fabrication disorder, which demonstrate that the enhancements in optomechanical coupling rate are not accompanied by prohibitive radiative loss. The average disorder-limited quality factor for our design is found to be approximately $3 \times 10^5$, which corresponds to a net enhancement of the single-photon readout rate by a factor of 2 compared to the the state-of-the-art.

%Improvement in figures of merit is robust to fabrication disorder. 

\section{Body}

In its most basic form, an optomechanical system can be understood by considering a Fabry-Perot cavity made of a fixed and a moving mirror. In this canonical picture, displacements in the moving mirror ($x$), results in a change in the overall length ($L$), and subsequently a change in the resonance frequency of the cavity ($ \Delta \omega_0 = \omega_0\frac{x}{L}$). The optomechanical interaction rate in this system is found to be proportional to the rate of change of the optical resonance frequency with respect to the mechanical displacement, which scales inversely with the cavity length $g \propto \frac{\partial\omega_0}{\partial x} = \frac{\omega_0}{L}$ \cite{Aspelmeyer:2014ce}. Previous work has relied on this principle for optimizing the optomechanical coupling by reducing the mode volume of an optical cavity, which in dielectric resonators can approach the diffraction limit ($V_{\text{eff}}\sim (\lambda/n)^3$, set by the effective in-material wavelength of light) \cite{Eichenfield:2009ev}.

While the diffraction limit is sometimes considered as the ultimate bound for mode confinement in an all-dielectric resonator, recent progress has established deep sub-wavelength confinement (with large intrinsic quality factors) by employing defects in photonic crystals waveguides ($V_{\text{eff}}\sim{10}^{-3}{(\lambda/n)}^3$ \cite{hu2018a}). Additionally, this concept has been extended to mechanical waves, where embedding defects in phononic crystals has resulted in ultra-small mechanical mode volumes ($V_{\text{eff}}\sim{10}^{-4}\lambda_{\textrm{mech}}^3$ \cite{schmidt2020}). The combination of these developments motivates exploring geometries for co-confinement of optical and mechanical waves to sub-wavelength dimensions, with a potential for enhanced optomechanical coupling. For this purpose, we investigate defects in photonic/phononic crystals in silicon nano-beams, where the material and structural properties result in comparable phase velocities for (GHz-frequency) mechanical and (telecommunication-band) optical waves.

 We begin by demonstrating sub-wavelength confinement of light in a narrow air gap that forms a defect in a one-dimensional photonic crystal waveguide. For designing the primitive unit cell, we start with the simple case of a circular hole in a silicon nanobeam, and successively modify the cell by introducing defects in form of a narrow beam, and a narrow beam interrupted by a gap (see Fig.~1a-c). For a structure with a lattice constant of $\Lambda$, we perform finite element method (FEM) simulations with periodic boundary conditions ($k_y \Lambda = \pi$, see Fig.~1a) to find the optical mode and calculate the mode volume following the definition
\begin{equation}
\label{eq:VE}
    V_{\mathrm{E}} =  \frac{\int dV\, \epsilon\mathbf({r}) |\mathbf{E(r)}|^2} {\text{max}\left[\epsilon\mathbf({r}) |\mathbf{E(r)}|^2\right]},
\end{equation}
where $\mathbf{E(r)}$ is the electric field profile and $\epsilon = \epsilon_0\epsilon_r$ is the electric permittivity. The reduction of mode volume in the structures with a defect can be understood by considering the electromagnetic boundary conditions. In the case of the narrow beam inside the circular hole, the continuity of the tangential component of the electric field results in the enhancement of the peak electromagnetic energy density by a factor of $\epsilon_r$. Similarly, for the case of the gap interrupted beam, the continuity of the normal component of the displacement field, $\mathbf{D(r)}$, on the dielectric/air interface (inside the gap) leads to further enhancement of the peak energy density. Repeated application of the strategies in these two steps results in a `bowtie' structure, which can be considered the limit of an infinite concatenation of air gaps and narrow beams \cite{hu2016a,choi2017a}. As seen in Fig.~1d, a bowtie structure interrupted by a narrow gap at its the center supports an optical mode confined to the sub-wavelength air volume between the two tips.

\begin{figure}[t!]
\centering\includegraphics[width=13cm]{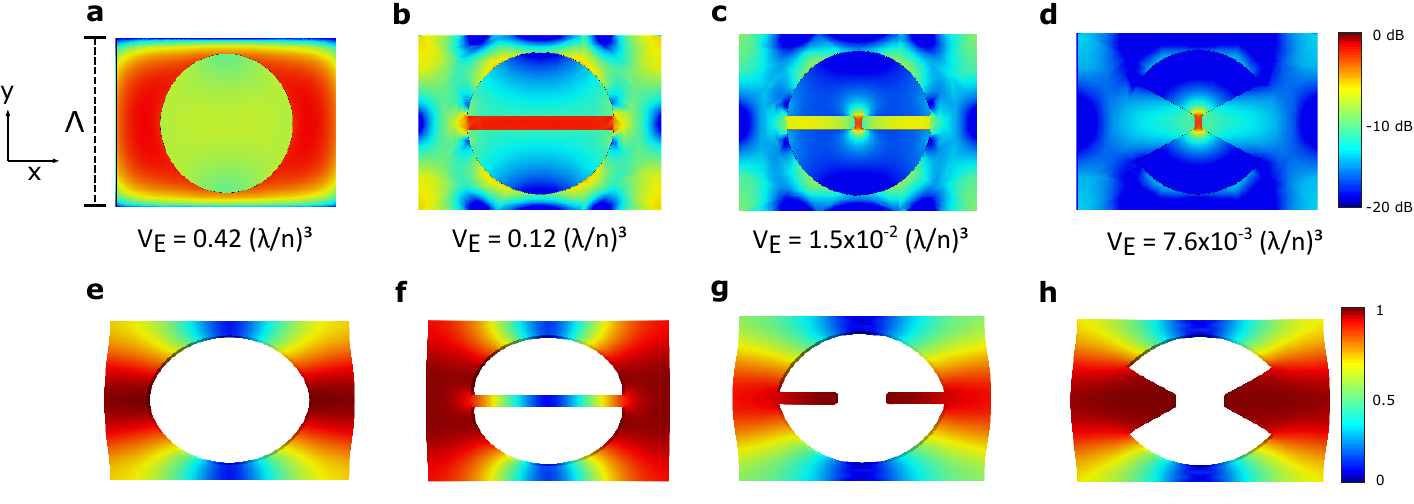}
\caption{(a-d) Simulated electric energy density in a photonic crystal unit cell. Wave propagation is along the y-direction, and the lattice constant is denoted by $\Lambda$. The quasi-TE mode has dominant electric field in the x-direction. The gradual improvement in the confinement with each step is evidenced by the reduction in the electrical mode volumes, expressed in terms of the wavelength in vacuum. The color map is in logarithmic scale. (e-h) Simulated displacement field for the `breathing' mechanical modes supported by the unit cell. The coloring depicts normalized mechanical displacement amplitude.}
\end{figure}

The confinement of the electromagnetic energy density in the air gap provides an opportunity to achieve coupling between the optical mode and the `breathing ' mechanical modes that typically arise in a nanobeam. Figure~1e-h, depicts the displacement profiles of these modes in our design (assuming periodic boundary conditions, $k_y \Lambda = 0$). The change in the effective size of the air gap caused by the mechanical motion gives rise to optomechanical coupling, with the single-photon coupling rate calculated as 
\begin{equation}
\label{eq:g_MB}
    g_{0} = \frac{-\omega_0}{2} \frac{\oint dA\, (\mathbf{Q(r)}\cdot \hat n) (\Delta\epsilon|\mathbf{E^\parallel(r)|^2} - \Delta\epsilon^{-1} |\mathbf{D^\perp(r)}|^2) }{\int dV \,\epsilon\mathbf({r}) |\mathbf{E(r)}|^2}.
\end{equation}
Here, the surface integral is evaluated over the dielectric-air interface, $\hat n$ denotes the unit surface normal, $\Delta\epsilon = \epsilon_0 (\epsilon_r-1)$, $\Delta\epsilon^{-1} = \frac{1}{\epsilon_0} (1/\epsilon_r-1)$, and $\mathbf{Q(r)}$ is the displacement profile of the mechanical mode. To calculate the single-photon optomechanical coupling, we have normalized the kinetic energy of the mechanical mode to the quantum zero-point fluctuation
\begin{equation}
\label{eq:x_zpf}
 \omega_{\mathrm{m}}^2 \int dV \rho(\mathbf{r}){ {|\mathbf{Q(r)}|}^2} = \frac{1}{2}\hbar \omega_{\mathrm{m}},
\end{equation}
where $\rho(\mathbf{r})$ and $\omega_{\mathrm{m}}$ are the local mass density and the mechanical resonance frequency.

\begin{figure}[b!]
\centering\includegraphics[width=12.5cm]{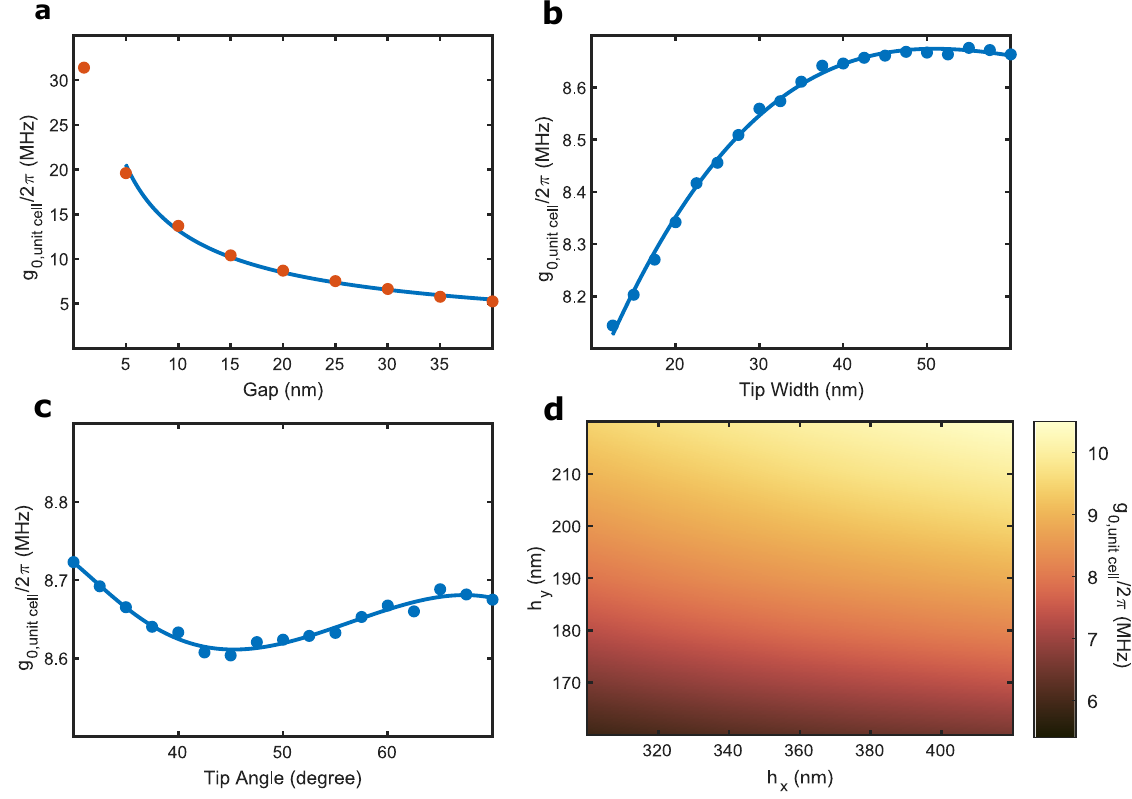}
\caption{ (a) Single-photon optomechanical coupling of the unit cell as a function of gap between the bowtie tips. The unit cell length is adjusted for each value of the gap to keep the frequency of the optical mode approximately constant. For the range 10-40 nm, we fit a power law of the form $\propto \text{gap}^{-\alpha}$, with $\alpha = 0.64$. The simulation at 1 nm is provided as a reference to present how coupling strength behaves at the ultimate limit of small gaps, which breaks the power law at larger gap values. (b) Dependence of coupling rate on the bowtie's tip width. (c) Variation in the unit cell coupling strength with the tip's angle.  (d) Dependence of the unit cell coupling rate to the geometry of the central hole in the unit cell. To widen the search space, the circular hole in the beam is replaced with an ellipse parameterized with two distinct diameters ($h_x$ and $h_y$).}
\end{figure}

The coupling rate from Eq.~\eqref{eq:g_MB} is strongly dependent on the difference in the electric energy density across the moving dielectric-air interfaces. In our design the mechanical mode has significant displacement near the bowtie tips (see Fig.~1h), where the optical energy is stored. The degree of localization of the mechanical field can be quantified by calculating the effective mode volume for the displacement field 
\begin{equation}
\label{eq:mech_vol}
 V_{\mathrm{Q}} = \frac{\int dV \rho(\mathbf{r}){ {|\mathbf{Q(r)}|}^2}}{\text{max}\left[\rho(\mathbf{r}){ {|\mathbf{Q(r)}|}^2}\right]}.
\end{equation}
For the simulated unit cell we find $V_{\mathrm{Q}} = 1.2\times{10}^{-3} \lambda_{\textrm{mech}}^3$. While the good overlap and the strong confinement of the optical and mechanical fields indicate the possibility of achieving large optomechanical coupling, an explicit evaluation requires forming a cavity from a linear array of the proposed unit cell. Before attempting to do this, we first calculate the coupling rate with the optical and mechanical fields normalized to a single cell. This calculation is useful considering the simplified case of a cavity composed of $N$ unit cells, where the coupling rate in the cavity is directly proportional to that of the unit cell via $ g_0 = g_{0,\text{unit cell}}/ \sqrt{N}$ (assuming a uniform optical and mechanical energy distributions across the cells). Thus, calculating $g_{0,\text{unit cell}}$ can serve as a crude proxy for $g_0$, which allows us to use it as a performance metric in optimizing the design at the single-cell level. This approach is particularly advantageous in speeding up FEM simulations, which tend to be quite expensive for structures with multiple cells due to the need for fine meshing in the vicinity of the bowtie.

We study the variations in the optomechanical coupling at the unit cell level as a function of the various geometrical parameters in our design. We find the gap size between the two tips in the bowtie as the most dominant parameter, resulting in a monotonic increase in the coupling rate (accompanied by a reduction in the optical mode volume) that can be fitted by a power law (see Fig.~3a). This sharp increase is a strong indication of the role of confinement in boosting the optomechanical interaction. The choice of gap dimension in the range of 5-40 nm is motivated by previous reports of fabrication of bowtie structures with 12 nm wide bridges \cite{hu2018a} and 20 nm air gaps \cite{10.1038/ncomms8915,10.1063/1.4989677, 10.1088/1361-6528/abdceb}. The second studied parameter for controlling the coupling rate is the tip width of the bowtie. We find that even though the optical mode volume decreases in structures with a narrower tip, the optomechanical coupling rate tends to be independent of the width for a wide range of parameters (see Fig.~3b). This can be understood considering a quasi-static electric field between the two tips in analogy to the field lines in a parallel-plate capacitor \cite{choi2017a}. Finally, the optomechanical coupling rate is found to be a weak function of changes in the tip angle and the hole's geometry (see Fig.~3c,d).

\begin{figure}[b!]
\centering\includegraphics[width=13cm]{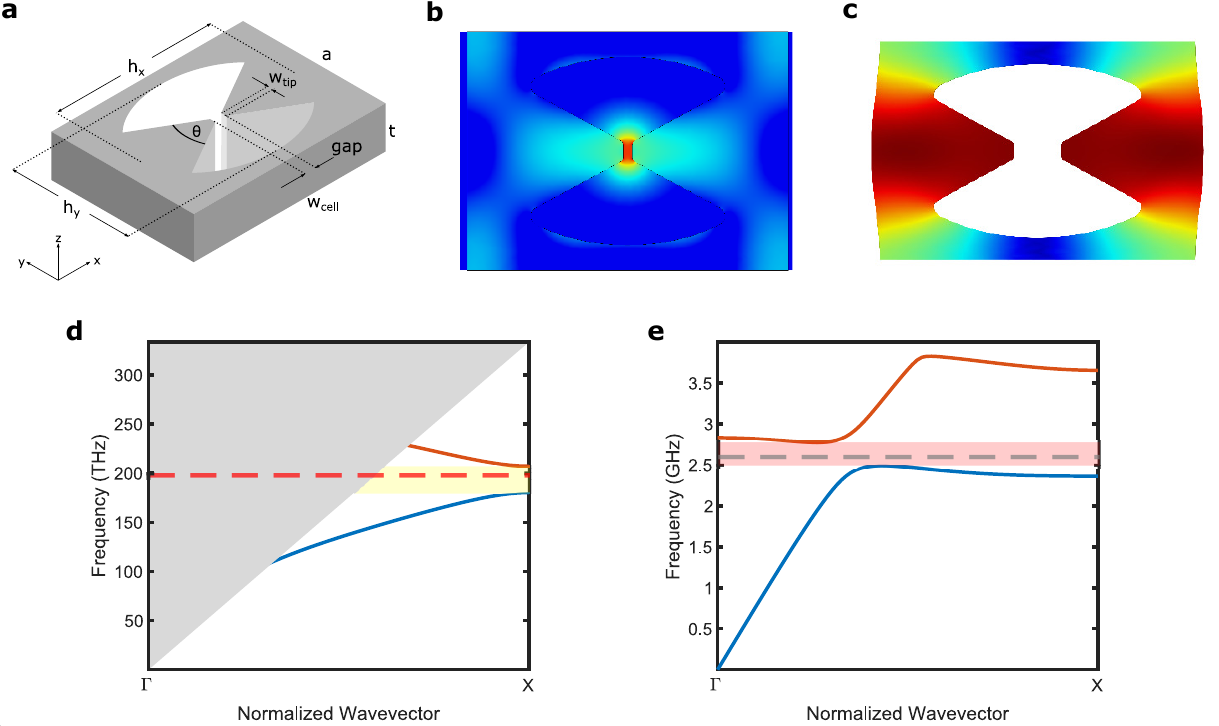}
\caption{(a) Isometric drawing of the unit cell. The nanobeam extends along the y-direction.  (b) Logarithmic electromagnetic energy density plot of the optical mode. The optical mode is at the X-point of the air band and has quasi-TE polarization with even x symmetry where the dominant electric field lies in the x-direction. (c) Displacement profile of the breathing mode at $\Gamma$-point with exaggerated deformation (d) Optical band structure of our unit cell, where only the quasi-TE bands are plotted. The red plot depicts the air band (mode of interest) and the blue band is the dielectric band. The grey shaded region is the light cone, and the yellow shaded region is our (quasi-)bandgap (for TE modes). (e) Mechanical band structure of the unit cell for the modes with even x and z symmetries. The breathing mode belongs to the $\Gamma$-point of the red band. The shaded region depicts the mechanical (quasi-)bandgap (for modes with even x and z symmetries).}
\end{figure}

Establishing the potential for enhanced coupling rate in the bowtie unit cell, we proceed to make a confined optomechanical cavity. The cavity is formed by modifying the geometry of the unit cells in a one-dimensional array to transition from `mirror' regions on the two extremities to a central cavity region. In this approach, excessive radiation loss can be avoided by constructing a cavity with smooth field envelopes based on a unit cell with optical and mechanical band gaps \cite{quan2011}. Figure~2 presents the mechanical and optical mode profiles and band structures for the primitive unit cell in our design. We employ elliptical holes in place of circular ones as controlling the ellipticity provides greater control over the mechanical properties. We find an optical band structure with a gap of about 27 THz centered around 1550 nm (see Fig.~2d). Our strongly localized optical mode corresponds to the X-point of the air band. Similarly, we find a gap of around 300 MHz, centered at 2.6 GHz, in the mechanical band structure (see Fig.~2e). The breathing mode of interest corresponds to the $\Gamma$ point of the upper band.

\begin{figure}[tbh!]
\centering\includegraphics[width=13cm]{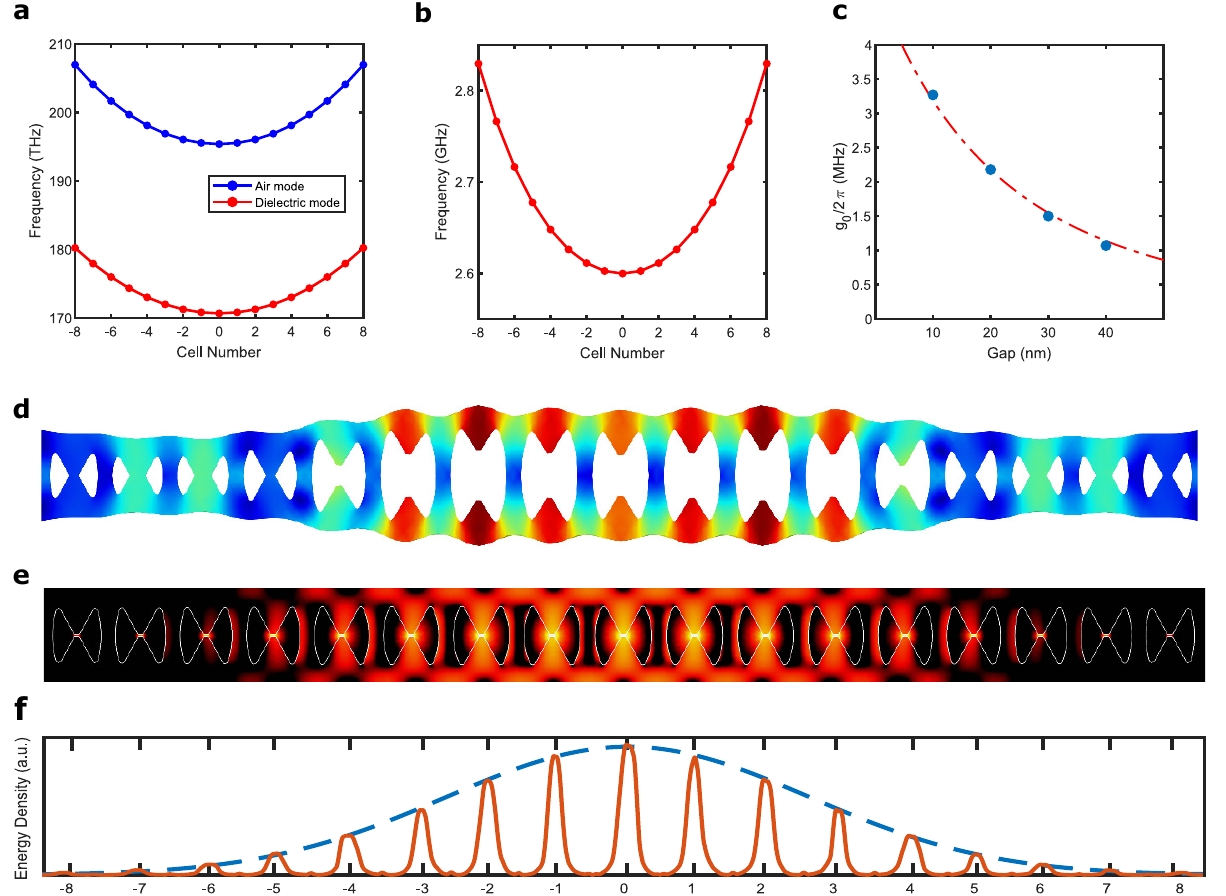}
\caption{(a) Frequency of the optical band at the X-point, as a function of the cell index in the central section of the cavity. (b) Frequency of the mechanical band at the $\Gamma$-point in the central section. (c) Single-photon optomechanical coupling rate for cavities designed based on unit cells with different gaps between bowtie tips. Dashed line depicts an extrapolated power-law fit. (d) Mechanical displacement profile with exaggerated deformation. (e) Logarithmic electromagnetic energy plot of the optical mode showing both localization and the envelope of the mode. (f) Electromagnetic energy density plotted along a line passing through the center of the nanobeam and extending through the nanobeam. Dashed line depicts a Gaussian fit to the envelope of the energy density in the cavity. For the cavity field profiles, only the central defect regions are visualized. The mirror cells outside the defect region (25 cells on each side) are absent from the figure. }
\end{figure}

In light of the information we have obtained from the unit cell parameter sweeps we proceed to build the optomechanical cavity. In order to confine the optical mode to the central region, we reduce the frequency of the optical air mode down into the bandgap by increasing the local lattice constants (see Fig.~4a). For the mechanical mode, we modify the frequency of the central cavity region by modifying the local ellipse parameters (see Fig.~4b). Quadratic tapering is used between the center unit cell and the mirror regions to obtain a smooth envelope. Using a primitive unit cell with a gap size of 20 nm and mirror sections with 25 cells (on each side), we find an optical mode at 1520 nm with a simulated radiation limited Q of $1.6\times10^6$ (see Fig.~4e) and with an optical mode volume of $6 \times 10^{-2} {(\lambda/n_{\mathrm{Si}})}^3$. The confined mechanical mode of the cavity is found at 2.58 GHz (see Fig.~4d). Evaluating the surface integral in Eq.~\eqref{eq:g_MB}, we find an optomechanical coupling rate of $g_0/2\pi\sim$2.2 MHz, which is a factor of 2 larger than the largest reported values in previous work on GHz mechanical frequency optomechanical crystals. We can further increase the optomechanical coupling to above $g_0/2\pi\sim$3 MHz by reducing the gap dimensions down to 10 nm (see Fig.~4d). Repeating the simulations for gap sizes in the range 10-40 nm, we find the radiation-limited optical quality factor to be consistently above a million.

The small radiation-limited loss in our design is encouraging considering the difficulty in attaining large quality factors in slotted photonic crystal cavities \cite{10.1364/oe.21.032468}. However, in a practical setting, the maximum attainable quality factor is further limited by the fabrication disorder \cite{minkov2013,minkov2014,lethomas2011}. We evaluate the performance of our design by modeling disorder as random displacements in the central position and changes to the lateral dimensions of the negative patterns etched into the dielectric structure \cite{minkov2013,maccabe2019,hagino2009}. These random changes can be modeled as independent and identically distributed Gaussian random variables with zero mean and standard deviation $\sigma$. We implement the disorder simulations variations in the parameters as follows; the central position of the holes vary by $\sigma$, the semi-axis lengths of the ellipses similarly vary by $\sigma$ and the gaps between tips vary by $2\sigma$. We note that the change in the gap also translates into a change in the tip width. We implement Monte-Carlo simulations to extract statistics for the optical mode in the presence of disorder.

Figure 5a presents the simulated decay rates as a function of disorder in our 20 nm gap design. Fitting the data, we find a linear relation between average decay rate and the variance of disorder ($\kappa \propto \sigma^2$), in line with previous findings in photonic crystal devices \cite{minkov2013,hagino2009}. We further bench-mark our deigns for a range of gap parameters using a disorder parameter of $\sigma = 1$ nm (based on previous reports in devices with a single-layer electron-beam lithography \cite{lethomas2011}). As evident in in Fig.~5b, we find an average quality factor of approximately $3\times{10^5}$ for the gap sizes of 20 nm and beyond. The disorder becomes much more significant for the 10 nm gap design, nonetheless the average quality factor remains in the vicinity of ${10^5}$. To put these results in a better context, we have simulated the effects of disorder on the design from Ref.~\cite{chan2012} which results in an average quality factors of $6\times10^5$.

\begin{figure}[tb!]
\centering\includegraphics[width=12cm]{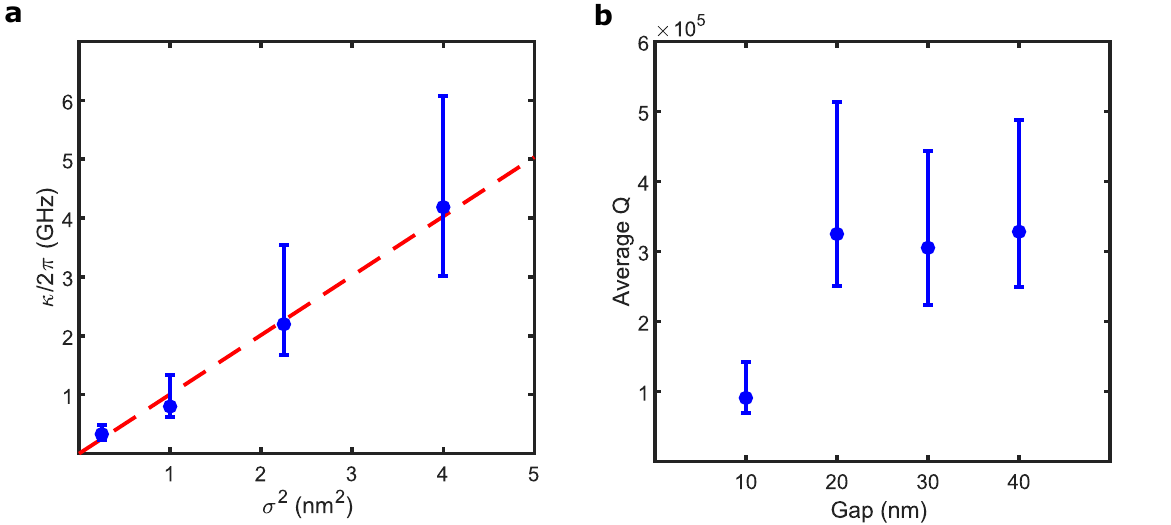}
\caption{(a) Dependence of the decay rate on the disorder present for 20 nm gap cavity. The fit crosses the origin and has a slope of approximately 1 GHz/$\text{nm}^2$. (b) Quality factor in the presence of 1 nm disorder for different gap values. Each data point consists of 20 realizations for both plots. The error bars represent one standard deviation.}
\end{figure}

\section{Conclusions and outlook}

In conclusion, we present a novel design concept, supported by numerical simulations, demonstrating a single-photon optomechanical coupling of $g_0/2\pi\sim$2.2 MHz and a radiation-limited quality factor of $1.6\times 10^{6}$ in a device with experimentally feasible parameters. These results are significant, considering the limited success of the past efforts in achieving large optomechanical coupling in devices with strong mode confinement \cite{schneider2016,davanco2012,grutter2015,safavi-naeini2010,li2010}, which despite the typical excess loss caused by the required small features have not been able to surpass the state-of-the-art limit of single-photon optomechanical coupling ($g_0/2\pi\sim$ 1 MHz \cite{chan2012}). We study the effects of fabrication disorder on the optical quality factor, and find a moderate reduction of Q (to $\sim 3\times10^{5}$) in our devices for realistic disorder levels. However, we note that the significantly larger optomechanical coupling rates in our design still translate to an overall improvement in the single-photon optomechanical readout rate $4g_o^2/\kappa$, and consequently larger phonon-photon conversion efficiencies compared to the state-of-the-art. The feasibility of device fabrication for the assumed small gap parameter ($\sim 20$nm), and the resulting limit on the maximum attainable optomechanical coupling rates remain to be studied in future work.

%\begin{figure}[tb!]
%\centering\includegraphics[width=11cm]{strain_figure_v2-eps-converted-to.pdf}
%\caption{(a) Logarithmic electric energy density of the quasi-TE optical mode at the X-point for a bowtie with %a dielectric bridge of 12nm width. The field is strongly confined in the bridge. (b) Logarithmic plot of the %strain field component ($|S_{xx}|$) that predominantly sets the the photo-elastic coupling for the optical mode %in (a). }
%\end{figure}

Looking ahead, we envision further progress in improving optomechanical coupling in devices with small mode volumes. While so far we considered contributions due to moving boundaries for achieving optomechanical coupling, contributions from the photo-elastic effect \cite{chan2012} provide an entirely separate route towards achieving this goal. Unlike the moving boundary contributions, which rely on the displacement profile, contributions from the photoelastic effect rely on the local strength of the strain field. We have investigated photoelastic coupling in a photonics crystal unit cell with an uninterrupted bowtie at its center, where the electrical and strain fields can be confined in a narrow bridge. Numerical modeling indicates a mode volume of $4\times10^{-3} (\lambda/n_{\mathrm{Si}})^3$ ($1\times10^{-5} \lambda_{\textrm{mech}}^3$) for the optical (strain) field in a structure with 12 nm wide bridge. However, we did not find a significant optomechanical coupling in this structure due to the challenges associated with getting strong overlap between the strain field and our air-band optical mode. Regardless, the possibility of co-confinement of the optical and mechanical modes to a shared sub-wavelength volume encourages future exploration of designs with superior optomechanical coupling.

%Finally, we note that the slotted photonic crystal cavities presented here can further find applications in high precision sensing of refractive index changes in extremely small sensing volumes due to the strong concentration of the electric energy in the air region\cite{jagerska2010}. Possible applications includes sensing the of gases \cite{jagerska2010} and concentration of liquid solutions \cite{xu2019}. For our optical cavity, we obtain a simulated sensitivity of  $ \Delta \lambda / \Delta n = 580 \,\text{nm}/\text{RIU}$ which is comparable to the state-of-the-art devices. 

\bibliography{OMC.bib, small_mode_OMC.bib}

\end{document}